\documentclass[preprint,prx,longbibliography]{revtex4-1}
\usepackage{subcaption}
\usepackage{graphicx}
\usepackage{graphics}
\usepackage{comment}
\usepackage{amsmath,amssymb,amsthm} 
\usepackage{scalerel,stackengine}
\usepackage{bm}
\usepackage{verbatim}
\usepackage{float}
\usepackage{xcolor}
\usepackage{color,soul}
\begin{document}
\title{ How to calculate temperature and energy in Molecular Dynamics }
\author{S\o ren Toxvaerd}
\email{ Submittet to J. Chem. Phys.; st@ruc.dk}

\affiliation{ Department of Science and Environment, Roskilde University, Post box 260, DK-4000 Roskilde, Denmark}

\begin{abstract}
Molecular Dynamics (MD) simulation of the dynamics of classical objects
has become incredibly popular in  natural sciences since the method was developed in the sixties. A conservative estimate is that it involves many hundreds
of thousands of scientific articles and all subdisciplines of natural sciences. The earliest MD
simulations also introduced a simple time-symmetric expression for the momentum and velocity,
which has since been used uncritically in almost all  MD simulations. The expression is incorrect 
 and can lead to significant errors. But it is straightforward to replace the expression with the correct one,
	and the motive of the article is to draw attention to how to calculate velocity, kinetic
energy,  temperature, and energy in an MD simulation correctly.
Newton's algorithm of the discrete dynamics
 has the same quality as his analytical dynamics, and Newton's discrete dynamics conserves the energy exactly. 
That is the second motive for the article, namely to
show that conservation of energy in a conservative system is only obtained by the correct expressions for the velocity and kinetic energy.
\end{abstract}
\maketitle

\section{Introduction}

Molecular Dynamics (MD) simulation of the dynamics of systems of classical mechanical objects
has become incredibly popular in the natural sciences since the method was developed in  the
sixties \cite{Rahman1964,Verlet1967}. A conservative estimate is that it involves many hundreds
of thousands of scientific articles and all subdisciplines of natural sciences \cite{arxiv}. The simulations
use algorithms for the time evolution, and almost all of the many MD simulations use the same
algorithm where a force impulse spontaneously changes an object's momentum.   The algorithm is formulated by Newton in his famous work \textit{Principia} \cite{Newton1687},  but has since been
reformulated and appears under different names. The dynamics is, however, independent
of which reformulation has been used in a simulation \cite{Toxvaerd2023,Toxvaerd2025}.

The earliest MD simulations also introduced a simple time-symmetric expression for the momentum and
velocity at the time when the force impulse acts,
which has since been used uncritically in almost all of the many MD simulations.
Verlet introduced the expression in 1967 \cite{Verlet1967}, obtained from a symmetric Taylor
expansion in time of the  positions \cite{note},
and in standard books for MD simulations it is presented the first leading therm in this expanion
 \cite{Tildesley,Frenkel}. 
However, as one can realize, the expression is systematic incorrect, 
 and it can lead to significant errors  \cite{vanGunsteren2007,Jung2018}. But it is straightforward to replace the incorrect expression
 with the correct one  \cite{Jung2018,Toxvaerd2024}, and
that is one of the motives for this article. Namely to draw attention to how the velocity, kinetic
energy,  temperature, and energy in an MD simulation with Newtons discrete algorithm are calculated correctly.

 \begin{figure}
          \begin{center}
         \includegraphics[width=5.6cm,angle=-90]{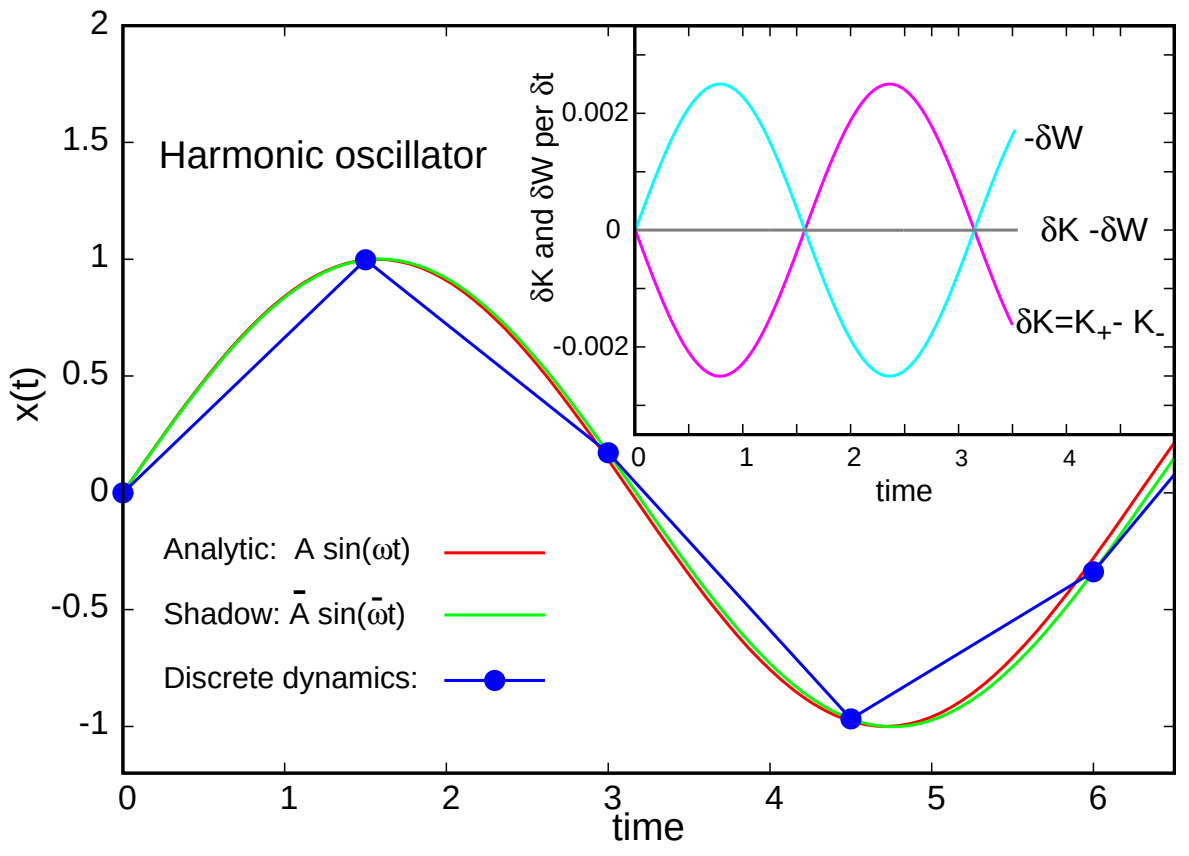}
		    \caption{ Dynamics for a harmonic oscillator. The discrete positions and the trajectory in blue are the
                  discrete dynamics for a harmonic potential, and the corresponding analytic dynamics $x(t)=A sin(\omega t)$ in red.
                  The trajectory $x(t)=\tilde{A} sin(\tilde{\omega} t)$  for the shadow Hamiltonian $\tilde{H}$ is
                  in green. The dynamics are for $A=\omega=1, x(0)=0$, and for  a very large time step $\delta t=1.5$.
                  The discrete positions are located on the analytic trajectory for
                  the shadow Hamiltonian. The inset is for $\delta t=0.005$, and it shows the change in kinetic energy $\delta K$,
		  per time step, Eq. (16) in magenta,
                  the work $W$ per time step, Eq. (15) in light blue, and the changes in energy per time step, Eq. (19), in gray.}
          \end{center}
  \end{figure}

Newton's formulation of discrete dynamics
is interesting because it has the same quality as his analytical dynamics.
It is time-symmetric and conserves the total momentum, angular momentum, and energy of a system \cite{Toxvaerd2023,Toxvaerd2025}. It can
further be shown that there probably exists a "Shadow Hamiltonian (SH)"   for the discrete Newtonian dynamics, corresponding to the Hamiltonian for analytic
dynamics, and where the discrete positions lie on the corresponding analytic trajectories  for SH \cite{Toxvaerd1994}.

The shadow Hamiltonian $\tilde{H}$ can be derived for a harmonic potential, and the energy of the discrete dynamics is conserved exactly per time step \cite{Toxvaerd1994}. The different dynamics for
the discrete versus the analytic dynamics are shown in Figure 1, and for a very large time increment $\delta t$, where the discrete dynamics jumps 
with only $\approx$ five positions per oscillation. For the dynamics shown in Figure 1 with
the analytic solution $x(t)=Asin(\omega t)$ with $A=\omega=1$ and $\delta t=1.5$ the energies are $E=0.5$ for the analytic dynamics
and $\tilde{E}=0.491062..$ for the analytic shadow Hamiltonian  $x(t)=\tilde{A}sin(\tilde{\omega} t)$
with $\tilde{A}= 1.001199..$ and $\tilde{\omega}= 0.989835..$  \cite{Toxvaerd2023}.
The inset is for $\delta t=0.005$, and it shows the change in kinetic energy $\delta K$,
		  per time step, Eq. (16) in magenta,
                  the work $W$ per time step, Eq. (15) in light blue, and the changes in energy per time step, Eq. (19), in gray.

  For other potentials and systems, one has only derived the first term in a functional expansion of the  energy for the discrete Newtonian dynamics,
and higher-order terms in the expansion are extremely complicated and have so far not been derived.
However, calculations of the energy of the discrete Newtonian dynamics, with the first term included, 
indicate a rapid convergence of the expansion toward the corresponding analytic energy \cite{Toxvaerd2012}.

 The existence of a shadow Hamiltonian means that there is no qualitative difference between the discrete Newtonian dynamics
 and a corresponding analytical dynamics. So the qualitative dynamical properties of discrete Newtonian dynamics versus traditional analytical dynamics
are the same. 
That is the second motive for the article, namely to show
that the energy is conserved exactly, consistent with the corresponding  conserved value for the  Shadow Hamiltonian,
but only with the correct expression for the kinetic energy and temperature.

\section{Discrete Newtonian Dynamics}
\subsection{Newtons discrete algorithm}
 Newton formulated the  MD algorithm for discret dynamics in $\textit{Proposition I}$  very first  in $Principia$. Below is an algebraic formulation.

Consider a classical conservative system of $N$ objects where the momentum and the velocity
\begin{equation}
	\textbf{v}_i(t -\delta t/2) \equiv \frac{\textbf{r}_i(t)-\textbf{r}_i(t-\delta t)}{\delta t} 
\end{equation}
for object No. $i$
at the position $\textbf{r}_i(t)$  and time $t$ 
  is changed by a discrete force-impulse $\textbf{f}_i(t)$ to
\begin{eqnarray}
	m_i \textbf{v}_i(t +\delta t/2)= \nonumber \\ 
	 m_i\frac{\textbf{r}_i(t+\delta t)-\textbf{r}_i(t)}{\delta t}
			=m_i\frac{\textbf{r}_i(t)-\textbf{r}_i(t-\delta t)}{\delta t} +\delta t \textbf{f}_i(t)	
 \end{eqnarray}
 according to
Newtons second law.
 The objects are again affected by force impulses
at time $t+\delta t$  after  a time increment $\delta t$ , etc.
By a rearangement of the terms in Eq. (2) one obtains Verlet's formulation of the dicrete algorithm.
\begin{equation}
	\textbf{r}_i(t+\delta t)=2 \textbf{r}_i(t+\delta t)- \textbf{r}_i(t) +\frac{\delta t^2}{m_i} \textbf{f}_i(t).
 \end{equation}
 The momenta, and the kinetic energy are constant inbetween the discrete force-impulses, but they change
 at the discrete times according to Eq. (2), and the kinetic energy changes from 
  \begin{equation}
	  K_-(t-\delta t/2)=  \sum_i^N \frac{1}{2}m_i \textbf{v}_i(t-\delta t/2)^2 
\end{equation}
to 
  \begin{equation}
	  K_+(t+\delta t/2)=  \sum_i^N \frac{1}{2}m_i \textbf{v}_i(t+\delta t/2)^2 
\end{equation}

 Verlet  also introduced the time symmetrical expression for the veloceties at time $t$
 \begin{equation}
	\textbf{v}_{0,i}(t)=\frac{\textbf{r}_i(t+\delta t) -\textbf{r}_i(t-\delta t)}{2\delta t},
\end{equation}
and in the many MD simulations one uses this expression for the veloceties at time $t$,
 to obtain the kinetic energy

 \begin{figure}
	  \begin{center}	  
	 \includegraphics[width=5.6cm,angle=-90]{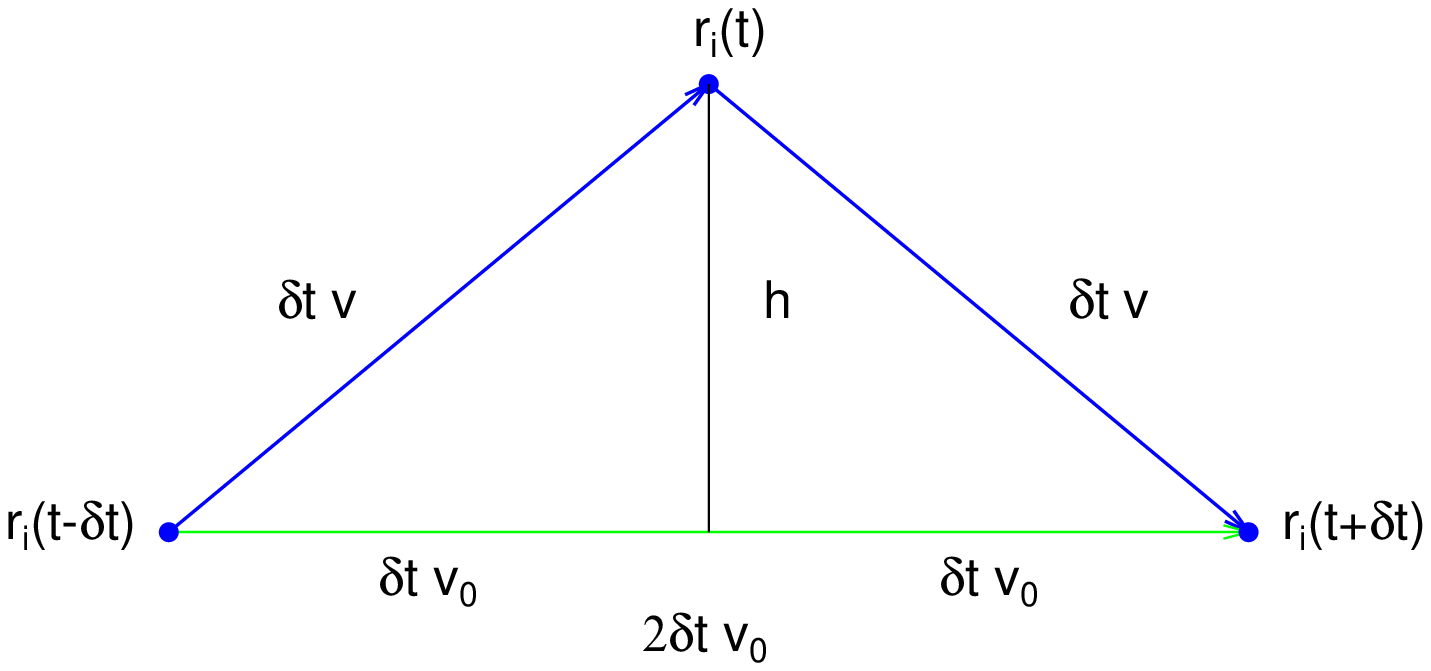}
		  \caption{ Illustrative example: In blue is  three positions and the trajectory at an elastic collision. Verlets approximation
		  for the length $2 \delta t v_0$ is in green.
		  The velocity before and after the collision
		  is $v$}, and the connection between $v$ and $v_0$ is $v_0=v\sqrt{1-(h/\delta t v)^2}<v.$ 
	  \end{center}		  
  \end{figure}

 \begin{equation}
	 K_0(t)=  \sum_i^N \frac{1}{2}m_i \textbf{v}_{0,i}(t)^2, 
\end{equation}
and the temperature
\begin{equation}
	T_0=\frac{<2 K_0(t)>}{N_f},
\end{equation}
for a system with $N_f$ degrees of freedom  in the system of $N$ objects.

 \textit{It is here the systematic errors in the temperature and kinetic energy are introduced.}

 The exact relation between the square of the veloceties in the time intervals $[t-\delta t,t]$ and 
 $[t,t+\delta t]$ and Verlet's expression at time $t$  from Eq. (6)  is drived in Appendix A. It is
\begin{equation}
	\textbf{v}_{0,i}(t)^2=\frac{1}{2}\textbf{v}_i(t+\delta t/2)^2+\frac{1}{2}\textbf{v}_i(t-\delta t/2)^2-
	\frac{1}{4}(\frac{\delta t}{ m_i }\textbf{f}_i(t))^2.
\end{equation}

 The correct expression for the kinetic energy and temperature at time $t$  is (see Section II B)
 \begin{eqnarray}
	 K(t)= \frac{1}{2} (K_-(t-\delta t/2)+K_+(t+ \delta t/2)),  \\
	 T(t)=\frac{<2 K(t)>}{N_f},
\end{eqnarray}
so Verlet's expression always underestimates the kinetic energy and the temperatures. The underestimations are usuallly
 of the order a few per cents or less for the traditional choise of state points and time increment $\delta  t$.

 The systematic
 deviation between $K(t), T$ and $K_0(t), T_0$  is not a consequence of the algorithm; it is due to Verlet's approximation. The
 systematic deviation is caused by the geometrical constraint,  the
 ``Triangle Inequality Theorem'', first 
 formulated by Euclid \cite{Euclid}.
 Three consecutive positions, which are not on a straight line (zero force) perform a triangle, and the length of the side
 opposite the central position is always less than the sum of the two other sides' lengths.
 But the exact relation Eq. (9) is only valid for Newton's algorithm, Eq. (2).

 An illustrative example is shown in Figure 2. The figure illustrates
 the systematic deviation between the velocety $v$ and $v_0$ for an elastic collision. 

The systematic deviation between the temperature  $ T$ and  $T_0(t)$    
 for the traditional Lennard-Jones (LJ)  system are shown in the next figure. Figure 3 shows the deviations  for $T=1$
 and for various state points and time increments. (Each values of  $T_0(t)$ are obtained from 
 two-three $NVE$ simulations for $N=2000$ LJ particles and for
	  $10^6$ time steps with $T\approx1$ by interpolation.) The deviations  between  $ T$ and  $T_0(t)$ increase for larger
 temperatures, density, and strength of the short-range repulsive forces.
 The effect of the strength of the repulsive forces is illustrated in the figure
 by showing the deviation between $T$ and $T_0$ for an $(n-6)$ LJ potential with
 the repulsive potential  between pairs of objects at short interparticle distances 
 proportional to $r^{-n}= r^{-18}$.

 \begin{figure}
	  \begin{center}	  
	 \includegraphics[width=5.6cm,angle=-90]{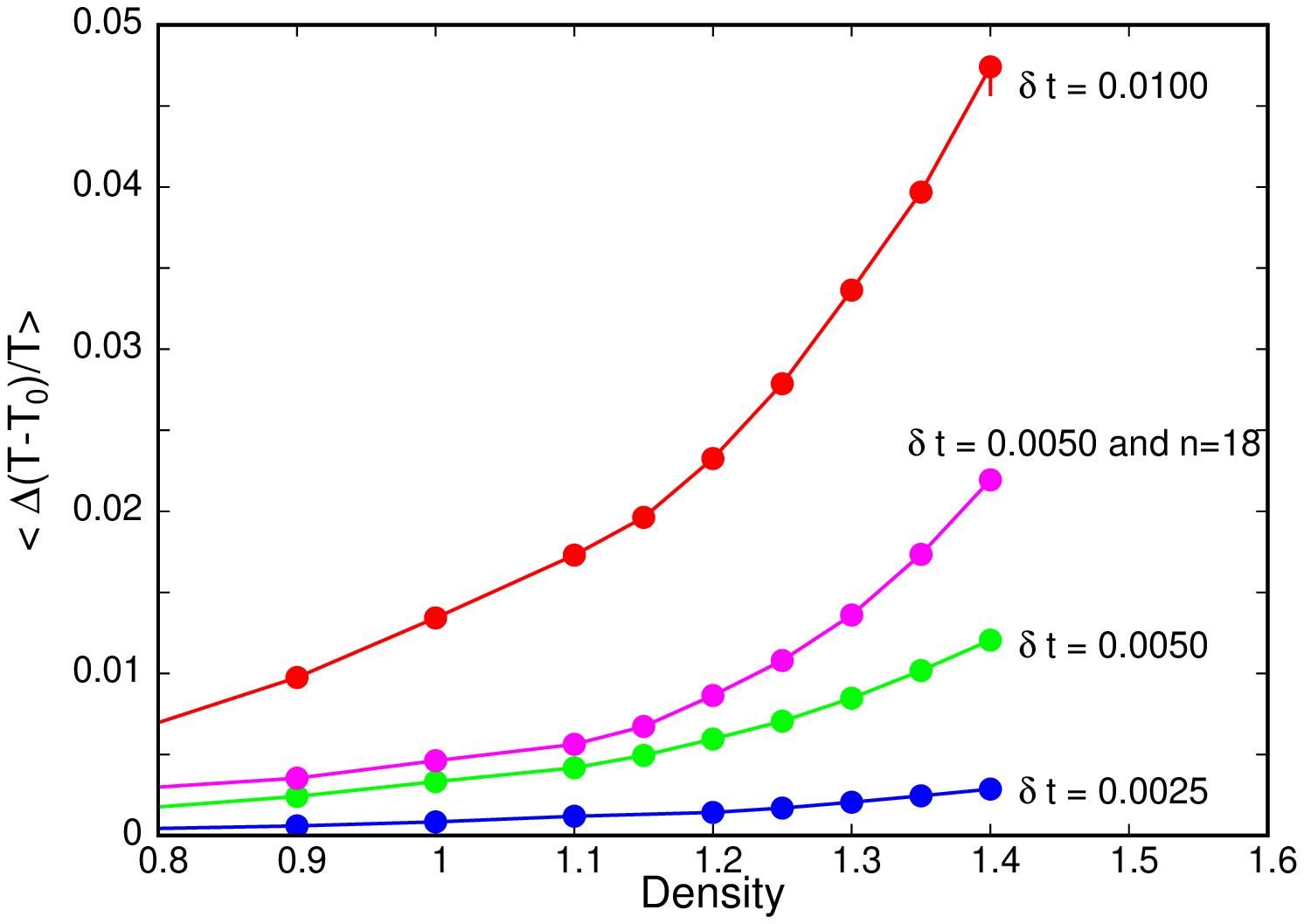}
		  \caption{ The relative difference $(T-T_0)/T$ between $T$, obtained from Eq. (10), and $T_0$ from Eq. (8), and for
		  differnt time increments $\delta t$, as a function of the density. }
  \end{center}		  
  \end{figure}

\subsection{ The connection between analytic and discrete Newtonian dynamics}

Classical mechanical dynamics is traditionally determined by Taylor expansions with polynomials in powers of the
discrete time step $\delta t$, where the highest power in  $\delta t$, indicates the order of the algorithm.
From this, Newton's discrete algorithm is a third-order predictor of a new position $\textbf{r}_i(t+\delta t) $ on the classical analytic trajectory.
However, Newton's discrete dynamics is exceptional compared with the dynamics obtained by all other algorithms for classical dynamics because its dynamics has the same qualities as the analytic classical dynamics.

The algorithm, Eq. (2), is symmetric with respect to time, and time-reversible and symplectic. Newton's third law, where forces 
$\textbf{f}_{ij}(t)=-\textbf{f}_{ji}(t)$ between pairs $i$ and $j$  cancles pairvise,   ensures conservation of momentum and angular momentum.
The conservation of energy can be determined in analogy with  the first law of thermodynamics  \cite{Toxvaerd2023}: 

\textit{
	In a system, the change in internal energy} $\Delta E$
\textit{
 equals the heat added Q
 minus the work done by the system,  W, expressed as}
\begin{equation}
	\Delta E=Q-W,
\end{equation}
and  with $\delta E=0$ for a conservativ system \cite{comment}.

Consider the time-interval $[t-\delta t/2, t+\delta t/2]$
and with positions $[((\textbf{r}_i(t)+(\textbf{r}_i(t-\delta t))/2,((\textbf{r}_i(t+\delta t)+(\textbf{r}_i(t)))/2]$,
and  force impulses $\textbf{f}_i(t)$. A force, $\textbf{f}(t)$,  performs a work on an object by
changing its momentum, and the work can be obtained from Eq. (2).
The work performed by the forces on the system of $N$ object in the time interval $[t-\delta t/2, t+\delta t/2]$ is
\begin{eqnarray}
	W= \sum_i^N \textbf{f}_i(t) \delta\textbf{r}_i \nonumber \\
	=\sum_i^N \textbf{f}_i(t) (\textbf{r}_i(t+\delta t/2)-\textbf{r}_i(t-\delta t/2))  \nonumber \\
	=\sum_i^N \frac{1}{2}\textbf{f}_i(t) (\textbf{r}_i(t+\delta t)-\textbf{r}_i(t-\delta t)). 
\end{eqnarray}
By rewriting  Eq. (3) to
\begin{equation}
	\textbf{r}_i(t+ \delta t) -\textbf{r}_i(t-\delta t))= 2(\textbf{r}_i(t) -\textbf{r}_i(t-\delta t))+\frac{\delta t^2}{m_i} \textbf{f}_i(t)
\end{equation}
and inserting in Eq. (13) one gets an expression for the work done by the forces in the time interval  $[t-\delta t/2, t+\delta t/2]$

\begin{equation}
	W=\sum_i^N  \textbf{f}_i(t) (\textbf{r}_i(t)-\textbf{r}_i(t - \delta t)) +\frac{\delta t^2}{2m_i} \textbf{f}_i(t)^2.
\end{equation}

The  heat added, $Q$, in the time interval $[t-\delta t/2,t+\delta t/2]$ is 
\begin{eqnarray}
Q= \delta K= 	K_+(t)-K_-(t)= \nonumber \\
\sum_i^N\frac{1}{2}m_i [\frac{\textbf(\textbf{r}_i(t+\delta t)-\textbf{r}_i(t))^2}{\delta t^2}-
\frac{\textbf(\textbf{r}_i(t)-\textbf{r}_i(t-\delta t))^2}{\delta t^2}].
\end{eqnarray},

By rewriting  Eq. (3) to
\begin{equation}
	\textbf{r}_i(t+ \delta t) -\textbf{r}_i(t)= \textbf{r}_i(t) -\textbf{r}_i(t-\delta t)+\frac{\delta t^2}{m_i} \textbf{f}_i(t)
\end{equation}
   and inserting the squared expression for  $\textbf{r}_i(t+\delta t) -\textbf{r}_i(t)$ in  Eq. (16), the change in kinetic energy
   in the time-interval   is
\begin{equation}
	Q=\delta K= \sum_i^N  \textbf{f}_i(t) (\textbf{r}_i(t)-\textbf{r}_i(t - \delta t)) +\frac{\delta t^2}{2m_i} \textbf{f}_i(t)^2,
\end{equation}
 and 
 with conservation of the energy 
\begin{equation}
	\Delta E= \sum_i^N \delta E_i= \sum_i^N  Q_i-W_i=0.
\end{equation}
So the expression for the kinetic energy, not by $K_0(t)$ but with $K(t)$ and with its change by the
discrete force impulses,  is consistent with the first law of thermodynamics and with energy conservation in a conservative system \cite{comment}. 

The energy conservation $\delta E_i=Q_i-W_i=0$ is for each object $i$, and at each discrete change of position for the system of $N$ objects, no matter what kind of
forces the object is exposed to. So the symmetry of Newton's discrete algorithm
and his second-order differential equation for analytic acceleration
ensures time reversibility, symplecticity, and energy conservation
for each object in the system, whereas the two other invariances, the conservation of momentum and
angular momentum in analytic and discrete classical mechanics are ensured by Newton's third law.

Moreover, there probably exists an analytic  dynamics (SH), where the discrete positions with Newton's algorithm are located on its analytic trajectories \cite{Toxvaerd1994}.
One can, as mentioned, only derive the first term in a functional expansion of the  energy,
and higher-order terms in the expansion are extremely complicated and have so far not been derived.
However, calculations of the energy for analytic dynamics, and with the first term included, 
indicate a rapid convergence of the expansion \cite{Toxvaerd2012}. Another indirect indication of
the existence of an SH has been performed by determining a high-order polynomial ($n=12$) fit to the discrete positions \cite{Ciccotti2026}. 
But unfortunately, the authors have used Verlet's wrong expression for the kinetic energy and temperature.

\section{Summary}
The many published articles on MD simulations use Newton's discrete algorithm, and most of the MD articles use Verlet's
expression for the velocity, Eq. (6),   to calculate the
kinetic energy and the temperature of the system. The temperature and energy are systematically wrong,
and the expression introduces a systematic error that can be significant.
The motivation for the article
is to draw attention to the fact that it is possible to calculate velocities, kinetic energy, and temperature consistently with the algorithm,
and in future simulations this error should be avoided.

Many of the articles with MD are for ``ensemble MD '', with constrained dynamics in order to obtain e.g. canonical $NVT$ dynamics. The
dynamics is constrained to sample  with a given input temprature.  Unfortunately most $NVT$ dynamics are with Verlets expression
for the temperature, i.e. $T_{input}=T_0$.  
These $NVT$ ensemble MD should also be corrected. An example of $NVT$ dynamics and with a Nos\'{e}-Hoover thermostate and with 
$T_{input}=T$ is given in \cite{Toxvaerd2024a},
but there are many other wayes to constrain the $NVT$ dynamics.

Newton's discrete dynamics is unique in that it has the same qualities as his analytical mechanics.
The energy is conserved, but only if one uses the correct expression, Eq. (10), for the kinetic energy.
There is probably a shadow Hamiltonian  where the discrete positions generated with the algorithm lie on an analytic trajectory,
but here too it is important to use the correct expression for the velocity and the kinetic energy,
if one compares the energy of a shadow Hamiltonian with the energy of the discrete Newtonian dynamics.

\section{Acknowledgment} Jeppe C. Dyre is gratfully acknowledged.

$   $ \\
$\textbf{Data Availability Statement}$ Data will be available on request.

\appendix
\section{ The exact realtion between the kinetic energies $K(t)$ and $K_0(t)$ }
The relations, Eqn. (9) and (10), between the squares of the velocities and the corresponding kinetic energies, were first derived by  J. Jung, C. Kobayashi, and Y. Sugita \cite{Jung2018}.
They obtained the relations by using the velocity Verlet algorithm (one of the  reformulations of Newton's discrete algorithm) and the
Trotter factorisation to derive the relations. They presented the relations as relations, which are correct to second order with respect to
the discrete time increment. It is correct when comparing kinetic temperatures and kinetic energies with the exact analytic solutions.
But the relations are  exact
relations for Newton's discrete dynamics, which is proved below.

The veloceties for each object $i$  at time $t$, where a discrete force impulse $\textbf{f}_i(t)$ acts, are  given by the discrete positions
$\textbf{r}_i(t-\delta t), \textbf{r}_i(t)$ and $\textbf{r}_i(t+\delta t)$. From the Verlet-formulation Eq. (3), one obtains
\begin{eqnarray}
	\textbf{r}_i(t+\delta t)-\textbf{r}_i(t-\delta t)= \nonumber \\
	2(\textbf{r}_i(t)-\textbf{r}_i(t-\delta t))+\delta t^2\textbf{f}_i(t)/m_i 
\end{eqnarray}
	and
\begin{eqnarray}
 \textbf{r}_i(t+\delta t)-\textbf{r}_i(t)= \nonumber \\
	\textbf{r}_i(t)-\textbf{r}_i(t-\delta t)+\delta t^2\textbf{f}_i(t)/m_i. 
\end{eqnarray}
By squaring the expressions Eqn. (A1) and (A2):
\begin{eqnarray}
	(\textbf{r}_i(t+\delta t)-\textbf{r}_i(t-\delta t))^2=4(\textbf{r}_i(t)-\textbf{r}_i(t-\delta t))^2+ \nonumber \\
	\delta t^4(\textbf{f}_i(t)/m_i)^2+4(\textbf{r}_i(t)-\textbf{r}_i(t-\delta t)) \delta t^2\textbf{f}_i(t)/m_i
\end{eqnarray}
and
\begin{eqnarray}
	(\textbf{r}_i(t+\delta t)-\textbf{r}_i(t))^2=
(\textbf{r}_i(t)-\textbf{r}_i(t-\delta t))^2+ \nonumber \\
	\delta t^4(\textbf{f}_i(t)/m_i)^2+2(\textbf{r}_i(t)-\textbf{r}_i(t-\delta t)) \delta t^2\textbf{f}_i(t)/m_i. 
\end{eqnarray}
By inserting $2(\textbf{r}_i(t+\delta t)-\textbf{r}_i(t))^2$ by Eq. (A4) in Eq. (A3) one gets
\begin{eqnarray}
	(\textbf{r}_i(t+\delta t)-\textbf{r}_i(t-\delta t))^2=2(\textbf{r}_i(t+\delta t)-\textbf{r}_i(t))^2+ \nonumber \\
	2(\textbf{r}_i(t)-\textbf{r}_i(t-\delta t))^2- \delta t^4(\textbf{f}(t)/m_i)^2,
\end{eqnarray}
or, by dividing Eq. (A5) by $(2\delta t)^2$,  one obtains the exact relation Eq. (9) 

 \begin{equation}
	\textbf{v}_{0,i}(t)^2=\frac{1}{2}\textbf{v}_i(t+\delta t/2)^2+\frac{1}{2}\textbf{v}_i(t-\delta t/2)^2-
	\frac{1}{4}(\frac{\delta t}{ m_i }\textbf{f}_i(t))^2
\end{equation}
 between the squares of the velocities.
	 The corresponding change in kinetic energy in a time interval $[t-\delta t/2, t-\delta t/2]$, centred at time $t$, and the  work obeyes
	 the first law of thermodynamics according to the proof in Section II B.

\end{document}